# Synthetic Participatory Planning of Shared Automated Electric Mobility Systems

**Jiangbo Yu (Correspondence)**

Department of Civil Engineering
McGill University
Montreal, QC H3A 0C3, Canada
Email: jiangbo.yu@mcgill.ca

**Graeme McKinley**

Department of Bioengineering
McGill University
Montreal, QC H3A 0C3, Canada
Email: graeme.mckinley@mail.mcgill.ca

**Abstract**

Unleashing the synergies among rapidly evolving mobility technologies in a multi-stakeholder setting presents unique challenges and opportunities for addressing urban transportation problems. This paper introduces a novel synthetic participatory method that critically leverages large language models (LLMs) to create digital avatars representing diverse stakeholders to plan shared automated electric mobility systems (SAEMS). These calibratable agents collaboratively identify objectives, envision and evaluate SAEMS alternatives, and strategize implementation under risks and constraints. The results of a Montreal case study indicate that a structured and parameterized workflow provides outputs with higher controllability and comprehensiveness on an SAEMS plan than that generated using a single LLM-enabled expert agent. Consequently, this approach provides a promising avenue for cost-efficiently improving the inclusivity and interpretability of multi-objective transportation planning, suggesting a paradigm shift in how we envision and strategize for sustainable transportation systems.

## Introduction

Mobility systems worldwide confront escalating challenges—aging infrastructure, increasing environmental impacts from transportation emissions, and widening service provision gaps that exacerbate social inequalities. Addressing these challenges demands smart and adaptive planning strategies to effectively leverage both mature and emerging technologies—including autonomous driving, vehicle electrification, low-latency communication, and Mobility-as-a-Service (MaaS) platforms. Shared Automated Electric Mobility Systems (SAEMS), exemplified by demand-responsive autonomous transit and passenger car services, autonomous electric micro-mobility systems, and unmanned aerial vehicle (UAV) delivery services, present a conceptual framework for integrating and leveraging these existing and promising technologies and addressing the escalating challenges.

However, the full advantages and potential side effects of SAEMS often remain uncertain due to environmental, technological, and socioeconomic factors. This ambiguity underscores the importance of integrating a broad spectrum of domain knowledge and perspectives—ranging from land use zoning to charging infrastructure engineering, and from local business operations to residents' daily experiences—into coherent planning processes. Therefore, an inclusive and anticipatory planning framework is crucial. It must assimilate diverse technical and non-technical stakeholder inputs to ensure the comprehensiveness and inclusiveness of proposed SAEMS plan and to ensure that the SAEMS develop and evolve sustainably and are aligned with broader societal goals.

Traditional participatory planning, while valuable, often faces various barriers that limit its effectiveness and inclusivity, including the logistical and administrative challenges of coordinating meetings, the financial and organizational obstacles of providing adequate incentives for participation, and difficulties in reaching consensus [1]. Longer participatory sessions can also lead to participant fatigue, resulting in lower-quality outcomes, and participants may churn or be unable to participate continuously due to other time commitments.

This paper introduces a synthetic participatory planning approach that leverages generative artificial intelligence to create digital avatars representing stakeholders in transportation systems planning and engineering. These avatars are powered by multimodal Large Language Models (LLMs) [2–5] and engage in virtual deliberations. As a significant revamp of an original human–AI collaborative engineering and planning framework and the embodied applications [6,7], the proposed method facilitates more effective and cost-efficient transportation planning processes than traditional methods, allowing for multiple participatory planning scenarios to refine system designs under various operational and strategic constraints.

Utilizing a structured and parameterized workflow, our approach employs LLM-enabled digital avatars to represent the knowledge and interests of potential stakeholders in the city of Montreal when planning SAEMS. This approach demonstrates LLMs' capabilities of aligning with stakeholders' needs and opinions with resultant SAEMS through specific performance metrics. By simulating the entire planning process—from identifying issues and setting objectives to conceiving and evaluating solution alternatives—our method offers a scalable, adaptable, and interpretable framework that combines the insights of both real and virtual stakeholders. The scalability of our framework lies in its ability to include numerous and varied stakeholder perspectives without requiring proportional increases in time and resources typically needed for in-person engagements. This adaptability allows the framework to be applied to different urban contexts and planning scenarios. Moreover, the interpretability is enhanced by the structured workflow and parameterized approach, which systematically documents the reasoning and decision-making process of both real and virtual stakeholders. By maintaining a clear and transparent record of how objectives are set, where alternatives are conceived and solutions are evaluated, the framework allows for easy traceability and validation of decisions, thus fostering trust and acceptance among all stakeholders involved.

To guide our investigation, we pose the following three research questions. How can digital avatars, calibrated with data from real stakeholders, enhance the inclusivity and effectiveness of the transportation planning process? In what way does the integration of digital and real stakeholders provide a scalable and adaptable framework for planning SAEMS? How does our structured and parameterized workflow improve the interpretability and controllability of planning outcomes compared to traditional methods?

The proposed method can be used to anticipate potential scenarios in the initial phases of planning or serve as the "digital twin" of the ongoing planning process. Figure 1 illustrates how this integration facilitates the use of diverse data sources to evaluate and optimize transportation solutions in a risk-free simulation environment. Through proposing the interactions between real stakeholders and their own or other stakeholders' digital counterparts, we offer a novel paradigm for developing interpretable and explainable AI-assisted decision support systems and contribute to human–AI teaming in transportation systems planning. We further highlight the main contributions as follows:

- **Synthetic Participatory Approach**: Introduction of a novel synthetic participatory method that leverages LLMs to create digital avatars representing diverse stakeholders in complex, multi-objective decision-making processes.
- **Innovative Use in Transportation Planning:** The proposed method demonstrates significant benefits, including scalability, adaptability, increased inclusiveness, and cost reduction compared to conventional participatory planning approaches.
- **New Paradigm for Human-AI Teaming**: This approach proposes a new paradigm for human-centered transportation systems planning and engineering, where human stakeholders and LLM-enabled digital avatars of stakeholders collaborate to create more sustainable and resilient infrastructure and communities.

The remainder of this paper is organized as follows. Section 2 offers a comprehensive review of the relevant literature, establishing the foundation for understanding the current landscape of SAEMS planning, participatory decision-making, and AI-enabled decision-support tools for transportation system planning. Section 3 details the methodology, focusing on the architecture and workflow of the synthetic participatory process involving digital avatars. Section 4 presents the context and settings of a case study in Montreal; this section specifies the background, prompting and output processing methods, baseline definition, and parameterizations for alternative scenario generations. Section 5 discusses the simulation results obtained from the case study, highlighting the effectiveness and implications of the synthetic participatory approach for SAEMS planning. Section 6 concludes the paper by summarizing the key findings, contributions, and implications for future research and practice in the field of SAEMS planning and synthetic participatory decision-making.

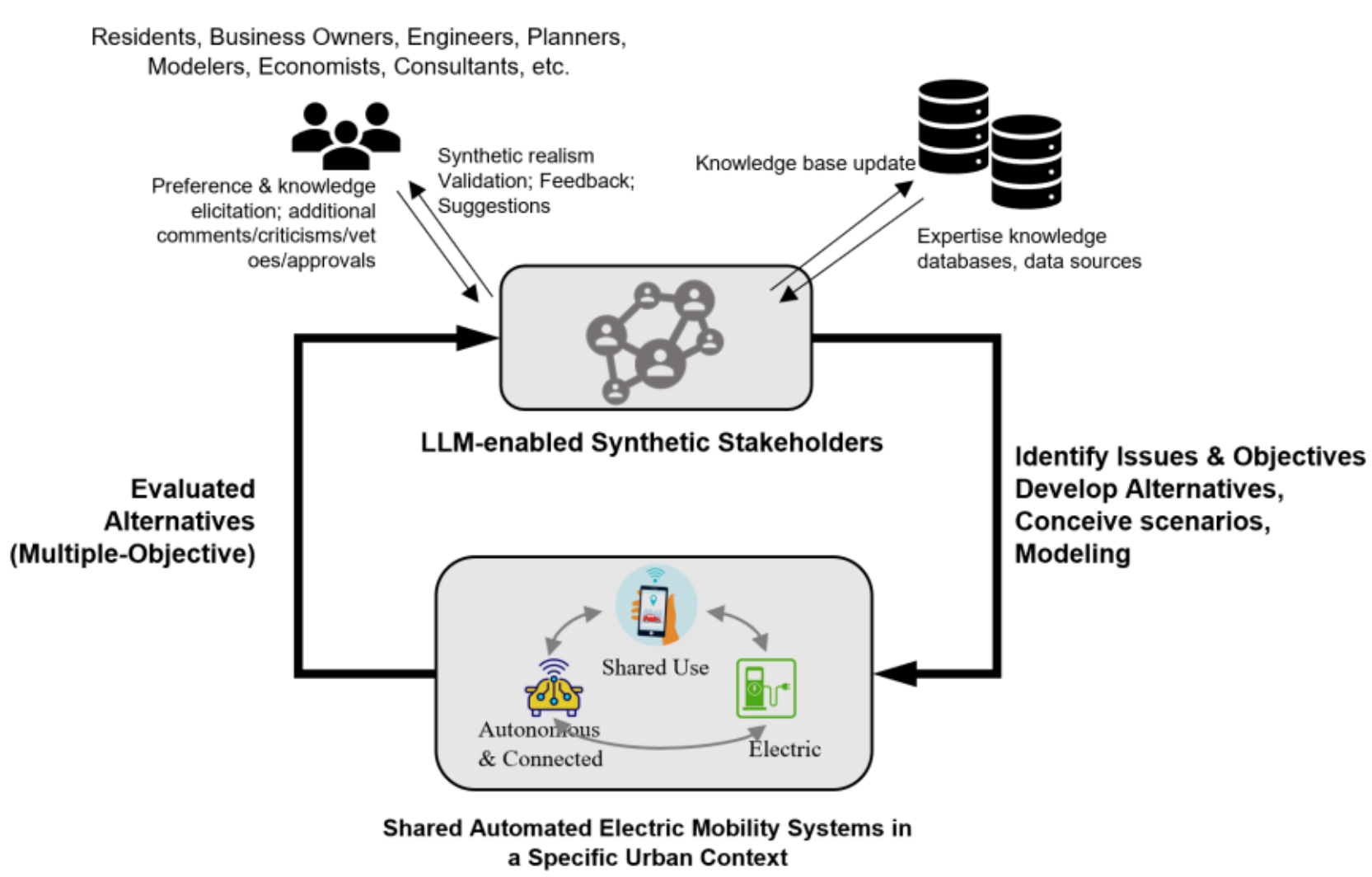

**Figure 1.** Conceptual illustration of the interaction between a group of stakeholders (and their digital representatives) and the transportation (sub-)systems of interest.

## Literature Review

The transition toward sustainable urban mobility is increasingly characterized by automation, electrification, and large-scale coordination through online digital platforms. The convergence of these technologies offers promising solutions to challenges such as inefficient road capacity utilization, high emissions, and availability constraints of human labor. The transformative potential of such systems has been underscored within the context of smart cities [8], highlighting the need for a holistic approach to urban mobility that leverages automation, connectivity, and electrification paradigms.

The planning and optimization of shared mobility systems requires the consideration of the integration of fleet management, curb management, depot siting and capacity planning, charging infrastructure, and user demand, some of which are highlighted by [15]. When the shared systems are on-demand or demand-responsive (rather than reservation-based), dynamically forecast future user requests to reposition empty vehicles efficiently becomes essential [16]. Ref. [17] assess the impact of charging infrastructure on the performance of shared autonomous electric vehicle services, providing insights into the operational considerations of shared mobility.

Although the individual technologies that enable automation, electrification, and shared mobility systems offer promising solutions to various urban mobility challenges, their combined effects on overall sustainability, resilience, and efficiency remain challenging. A joint optimization scheme has been proposed for the planning and operations of shared autonomous electric vehicle (SAEV) fleets providing mobility on demand [18]: a critical step toward multimodal, multi-purpose SAEMS. Researchers also proposed a comprehensive system design approach that incorporates new, emerging features of autonomous connected electric vehicle-based car-sharing systems [19]. This design also addresses the limitations of conventional car-sharing policies and electric vehicle technologies, such as costly vehicle relocation, restricted vehicle range, and lengthy recharge times. During operations, there is potential constructive coordination between charging activities and repositioning activities of operating SAEVs for reducing rider wait time, empty travel distance, and non-revenue operation time [20]. In addition to serving regular travel demand in business-as-usual conditions, novel use of surface and aerial SAEVs for disaster relief and community resilience has also been proposed [21]. In scenarios such as hurricane-induced power outages, pandemics affecting vulnerable populations, and earthquake-damaged infrastructure, SAEVs can be deployed to evacuate and rescue at-risk populations, provide essential supplies and services, and transport repair crews and equipment.

Despite the emphasis on technical aspects of SAEMS planning, there is a notable gap in integrating social considerations, particularly through participatory approaches. The integration of various constituent technologies within SAEMS underscores the complexity of urban transportation planning in the era of rapid technological change. A balanced approach that addresses both the technical and social dimensions of planning is essential for creating sustainable, efficient, and inclusive urban mobility solutions.

Participatory approaches in transportation planning are increasingly recognized for their ability to engage diverse stakeholders in the decision-making process, leading to more inclusive and sustainable urban mobility solutions. There is a wide range of participatory formats, ranging from the very structured Delph method [22,23] to highly free-style discussions. Researchers have examined the potential of participatory processes for simulation model development [24]. Multi-stakeholder decision-making problems are often multi-objective and multi-criteria in nature. Researchers have evaluated innovative ideas for public transport proposed by citizens using Multi-Criteria Decision Analysis (MCDA) [30]. In their study, the Analytic Hierarchy Process method ranks ideas based on feasibility, utility, and innovativeness. The results highlight the importance of citizen participation in generating and prioritizing innovative solutions for public transport, demonstrating the value of participatory techniques in enhancing the attractiveness of public transport systems. Researchers have also explored public opinions on personal mobility vehicle use in Palermo, Italy [31]. Their study underscores the significance of participatory planning processes in understanding and addressing the real needs of road users, emphasizing the role of sociodemographic characteristics in shaping public opinion and transport planning. Ref. [32] advocate for a transdisciplinary research and stakeholder engagement framework to address consumption-based emissions and their impacts on cities.

Role-playing represents a cutting-edge application of LLMs that extends beyond traditional uses, venturing into dynamic simulations of human-like behaviors and decision-making processes. This approach leverages the advanced capabilities of LLMs to create interactive scenarios where artificial agents simulate individual cognitive processes and complex social interactions. Ref. [3] discuss the risks associated with anthropomorphism in LLM role-playing, emphasizing the necessity of maintaining clear boundaries between human-like and machine-driven behaviors. They argue for the importance of ensuring that the outputs and actions of LLMs are strictly interpretable and aligned with their intended applications, avoiding misrepresentations that could lead to incorrect or unethical uses of the technology. Ref. [40] introduce the concept of generative agents that simulate human behavior within interactive environments. These agents are designed to perform daily activities, engage in social interactions, and respond to environmental cues, much like humans do. Ref. [41] explore the integration of LLMs in generative agent-based modeling, which is instrumental in mimicking human decision-making and interactions within social systems. LLM role-playing is an emerging field that offers significant potential across various domains; however, as this technology progresses, it is crucial to address ethical considerations and ensure the accuracy and appropriateness of the simulations to avoid potential misuse and societal harm.

## 3. Methodology

This section first introduces a synthetic participatory planning framework that integrates generative AI, particularly LLMs, for conceiving and planning SAEMS in the context of overall urban systems. The framework is grounded in the principles of participatory modeling and decision-making, the capabilities of generative AI for simulating complex human interactions, and the domain knowledge of transportation systems engineering and planning. We then present the workflow for utilizing such a framework in Sections 3.2–3.5.

### *3.1. Framework*

The proposed decision support system consists of three main components: a set of digital avatars representing stakeholders that form a synthetic team, a multimodal LLM for generating and simulating participatory planning processes and outcomes, and a computational and data storage platform that enables participatory modeling and evaluates design and planning alternatives. The LLM enables the intelligence of the digital representatives and simulates their activities and interactions to identify problems and solution alternatives, and then collaboratively develop models and evaluate these alternatives. Figure 2 shows the procedure of how the decision support system functions; the three components interact in a tri-phase workflow that allows iterations. Between different phases, the solid grey arrows represent the phase sequence, while the grey arrows with dotted outlines represent the potential rollback to the previous phase. Within each phase, the solid arrows represent the sequence of the internal steps, while the dotted arrows represent the potential rollback to the previous step. In the following subsections, we will present the details of each phase.

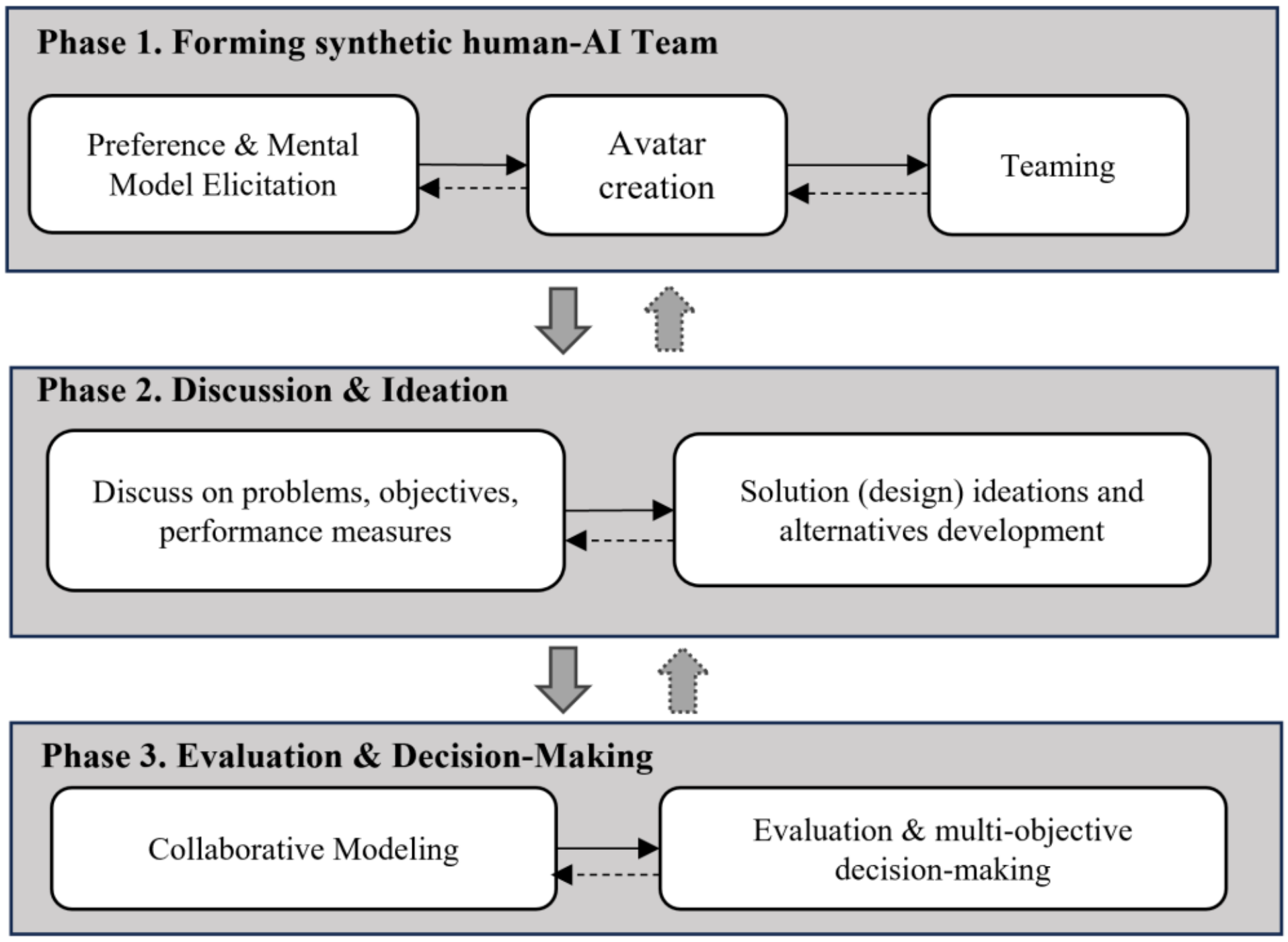

**Figure 2.** Synthetic participatory planning process of future mobility systems.

*3.2. Forming Synthetic Team (Phase 1)*

The initial phase starts with constructing the system prompting template $\rho_0 \in \mathrm{P}$ and scenario-specific parameters $\theta_0 \in \Theta$. $\rho_0$ and $\theta_0$ form the initial system prompting tuple $x_0 = \big(u_0(\rho_0, \theta_0)\big)$. Once $u_0$ is given, we can send $u_0$, along with the initial input $u_1$, to the LLM under use, which in turn produces $y_1$. We can represent this process as:

$$y_1 = \mathcal{LM}(u_1 | u_0) \tag{1}$$

where $u_0 = u_0(\rho_0, \theta_0)$ and $u_1 = u_1(\rho_1, \theta_1)$. Note that $\rho_0$ is a string with a variable length (or a fixed maximum length). We use $\mathcal{F}_\psi$ to denote the corresponding function that maps the input to the created set of digital avatars, which is:

$$\psi = \mathcal{F}_\psi(y_1) \tag{2}$$

An inverse problem of the process is to find the optimal $x_0$ and $u_1$ so that $\psi$ contains a list of digital avatar profiles with desired behaviors. We can also further add information about the team formation (e.g., a specific digital avatar will be the group facilitator or the moderator of the participatory planning session) by adjusting $\rho_0$ and $\rho_1$. We leave the study of an optimal calibration process to future research. We define the resultant output from this phase as a tuple of all the inputs and outputs (with sequence preserved). That is, $x_1 = (u_0, u_1, y_1)$.

*3.3. Collaborative Visioning, Ideation, and Design under Constraints (Phase 2)*

This phase involves utilizing the synthetic team to identify existing problems of the mobility systems, specifying planning objectives and the associated performance metrics (including their decision weights), and generating solutions (conceiving design alternatives and the associated implementation plans). Overall, we can denote the input–output relationship mapped by an LLM as:

$$y_2 = \mathcal{LM}\big(u_2\big|(u_0, u_1, y_1)\big) = \mathcal{LM}(u_2|x_1) \quad (3)$$

where $u_2 = u_2(\rho_2, \theta_2)$ and $\rho_2$ is a string of variable length that contains the background information about this stage of the decision-making process. Vector $\theta_2$ provides the parameters in this stage that can be joint with the string $\rho_2$ to form a complete string ($\rho_2$ has at least dim ($\theta_2$) blanks).

This phase starts with identifying the status quo, problems, objectives, and performance measures. This phase involves the synthetic participants identifying existing conditions, problems (including technical and financial constraints), objectives, and performance metrics, which form the basis for subsequent discussions and decision-making.

The prompt in this phase needs to specify the soft and hard constraints that form a clear decision space, so that $\wp = \mathcal{F}_{\wp}(y_2)$, $card(\wp) = M > 0$, and $\wp = B$, where $B$ is the complete set of $M$ constraints. Similarly, we can elicit objectives and performance metrics, $\mathcal{O} = \mathcal{F}_{\mathcal{O}}(y_2), card(\mathcal{O}) = V > 0$, and $\mathcal{K} = \mathcal{F}_{\mathcal{K}}(y_2), card(\mathcal{K}) = K \geq V$. Each element in $\mathcal{K}$ is a two-dimensional tuple that contains the score of the performance metric and the corresponding decision weight. That is, the $k$th element of $\mathcal{K}^k$ is $(g_k, w_k)$, $\forall k \in \{1,2, \dots, K\}$.

The present phase then generates solutions and develops alternatives using the identified status quo and the specified objectives and performance metrics. In this phase, participants ideate solution alternatives while considering constraints and practicality. The process encourages creativity while ensuring that proposed solutions are feasible and aligned with the identified objectives. This can be specified as $\mathcal{D} = \mathcal{F}_{\mathcal{D}}(y_2)$. The final output from this phase is a tuple $x_2 = (u_0, u_1, y_1, u_2, y_2)$.

*3.4. Collaborative Evaluation of Alternatives and Continuous Improvement (Phase 3)*

Evaluating design alternatives with specific decision contexts requires evaluation models. We can directly utilize the mental models captured by the profile of each stakeholder (or their digital avatar) to evaluate alternatives in multiple rounds of participatory modeling sessions and build a common ground, though we can also instruct more specific evaluation models. The prompts need to be specific so that the model has the capability to evaluate all the design alternatives previously defined with outputs consistent and coherent with all the performance metrics. Insights from avatar interactions enrich these models, enabling a dynamic approach to planning that incorporates diverse perspectives and preferences.

Once we have determined an evaluation method that allows multiple stakeholders to perform multi-criteria evaluation and decision-making, the third phase of the prompt ensures that the selected alternatives are evaluated based on various criteria, reflecting the interests and concerns of different stakeholders:

$$y_3 = \mathcal{LM}(u_3|x_2) \quad (4)$$

where $u_3 = u_3(\rho_3, \theta_3)$. We can then obtain the final evaluation score vector $E$ using a function (operator) $\mathcal{F}_E$ that converts $y_3$ to obtain the attribute value vector, $g$, and the corresponding decision weight vector, $w$, for the $K$ attributes. The results are $v_i = \sum g_{ik} \cdot w_k, i \in \mathbb{I}$. If $v_i \geq v_j$, option $i$ is the same or more preferable than option $j$. We can also denote it as $i \succcurlyeq j, i, j \in \mathbb{I}$. If $v_i < v_j$, option $j$ is more preferable than option $i$, or simply $i \prec j, i, j \in \mathbb{I}$. Combining the previous information along with the new outputs, we have $x_3 = (u_0, u_1, y_1, u_2, y_2, u_3, y_3)$.

Figure 3 summarizes the proposed three-phase process, where the inputs composed of strings and parameters feed into an LLM for obtaining the output responses. Then, all the prior inputs, outputs, and the new prompt (also composed of strings and parameters) are combined as the new input for the next phase (note that each phase might be composed of one or more prompting stages). In terms of continuous improvement, we can run a real participatory process and compare the results with the statistical average condition from the simulated participatory processes. The proposal is scalable in more than three phases. It is also allowable to decompose a phase into multiple phases. However, it is important to note that more phases lead to longer string $x$, which often requires higher token consumption (a measure of the usage of a commercialized LLM) and longer runtime.

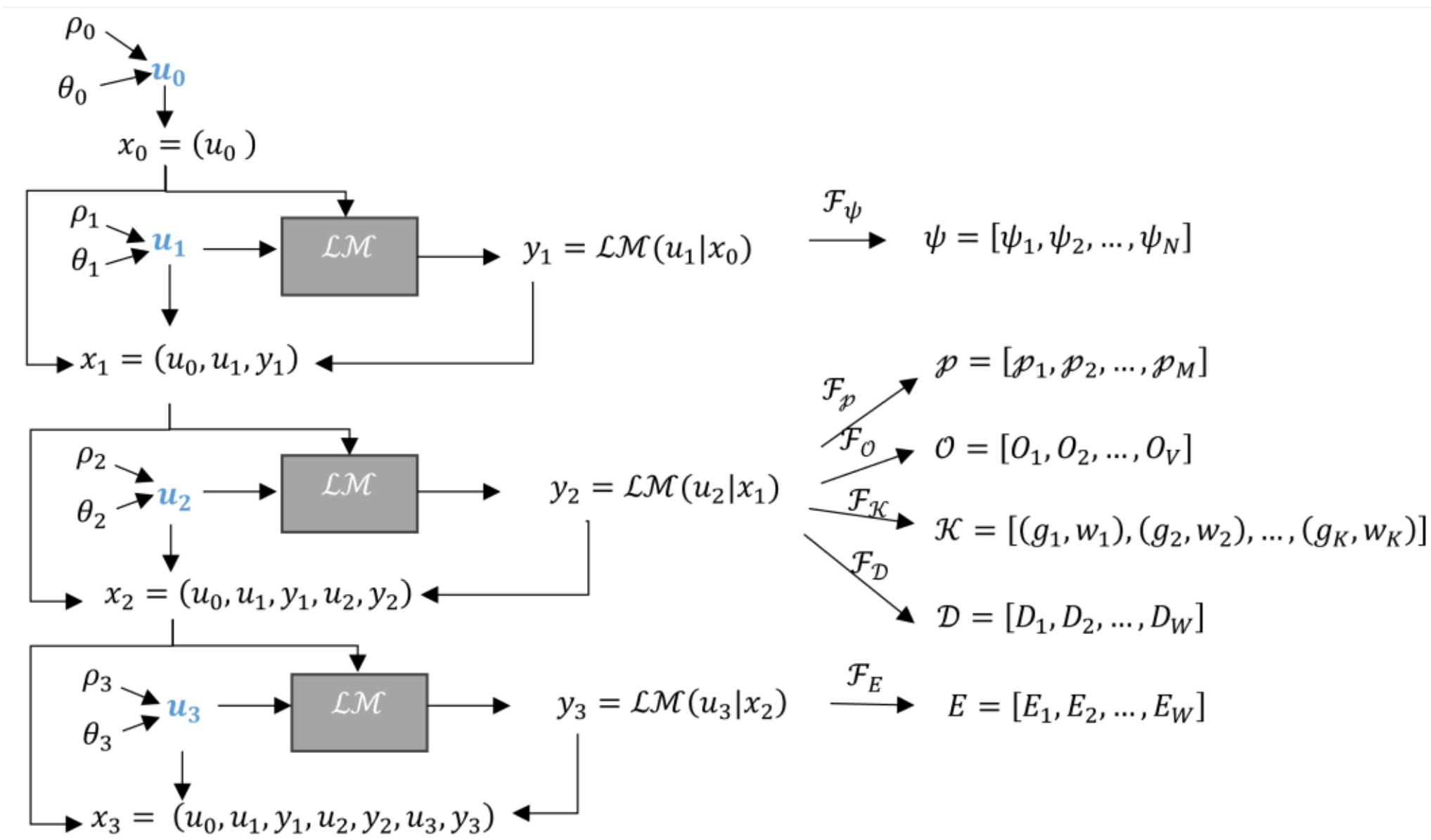

**Figure 3.** Illustration of the proposed multi-phase simulation procedure of the synthetic participatory approach for planning SAEMS.

In our methodology, AI assistants, potentially augmented by external knowledge databases (such as official design guides and factsheets), enabled by the LLM can also play a pivotal role in facilitating different forms of participatory engagement, tailored to the specific needs of the discussion format. In open discussions, AI assistants can function as moderators or facilitators, guiding the conversation flow and ensuring that all stakeholders' perspectives are considered. In Delphi sessions, these AI tools are utilized as summary generators, synthesizing complex inputs into concise overviews that aid in progressing rounds of expert consultation. Our framework is versatile, accommodating various configurations from a single AI assistant to multiple agents. In scenarios where a single AI assistant simulates a discussion or Delphi process, the attention mechanism of the LLM is crucial for simulating multi-agent cases. It selectively focuses on the profiles of each virtual agent from a new input prompt, drawing from historical outputs to simulate nuanced interactions. In multi-agent setups, we still need a "master" LLM (the focus of the present paper) to coordinate the inputs and outputs from different "child" LLMs.

### *3.5. Prompt and Parameter Sensitivity Analysis*

The sensitivity of the LLM outputs to inputs and parameters is analyzed by varying the prompts, or wordings in the prompts. This analysis is crucial for understanding how different linguistic formulations affect the calibration process (and hence, the output of a synthetic participatory planning process). Let $\theta = \{\theta_0, \theta_1, \dots, \theta_i, \dots, \theta_I\}, \theta \in \Theta$, be the set of $I + 1$ LLM parameter vectors for their corresponding prompts. The marginal sensitivity of $u_i$ to $\rho_i$ and $\theta_i$ can be presented as $\frac{\partial u_i}{\partial \theta_i}$ and $\frac{\partial u_i}{\partial \rho_i}$, respectively. After obtaining the initial vector $x_0$, we can evaluate the sensitivity of each parameter vector $\theta_i$, $i \geq 1$, by measuring the impact of varying $\theta_i$ on the output $y_i$. The expectation of the impact of a marginal variation of $\theta_i$, or simply $\delta_{\theta_i:y_j}$, can be represented as:

$$\delta_{\theta_i:y_i} = \mathbb{E}\left(\sigma_i \cdot \frac{\partial \mathcal{LM}(u_i|x_{i-1})}{\partial u_i} \cdot \frac{\partial u_i}{\partial \theta_i}\right) \quad (5)$$

where $i \geq 1$. Similarly, we can obtain the expectation of the impact of a marginal variation of $\rho_i$, or simply $\delta_{\rho_i:y_i}$. A high sensitivity indicates that the function is highly responsive to changes in the corresponding input or parameter, which may require careful control or robust design. On the other hand, a low sensitivity suggests that the function is relatively insensitive, allowing for more flexibility in design and operation. As not all the directions of perturbation are feasible (e.g., changing one letter or word in a prompt sentence might lead to an incomprehensible or grammatically incorrect sentence), the equation above might not be applicable, not to mention that the differentials might not exist. Therefore, we have $\sigma_i$ being a dummy variable—when the direction along $\theta_i$ is both differentiable for $u_i$, we have $\sigma_i = 1$; otherwise, zero. We can continue this evaluation process to obtain the impact of varying $\theta_i$ on $y_j$, where $j > i$.

Due to the complexity of the internal mechanism of an LLM, even if we have complete information about the model architecture and parameters, it might still be impractical to obtain the analytical solution to the sensitivities. However, the formulations above still provide valuable insights on how the sensitivity of prompting information on the output should be evaluated.

## 4. Case Study

This case study delves into the application of the synthetic participatory approach within the urban mobility planning landscape of Montreal—a city renowned for its dedication to sustainable transportation and pioneering infrastructure initiatives. The goal of this study is to highlight the enhanced efficacy of integrating LLM-enabled digital avatars with multi-criteria decision-making strategies into SAEMS planning.

### *4.1. Background*

Montreal, a major city in Quebec, Canada, is characterized by its diverse population, extensive public transportation network, and various social events and tourism activities. The city has been implementing sustainable mobility solutions, such as expanding its bike-sharing program, developing pedestrian-friendly streets, and introducing electric buses to its public transit fleet [42–44]. The Réseau Express Métropolitain (REM) [45] is an automated (driverless), electric light metro rapid transit system that is partially operational and is expected to be fully complete in 2027. Despite these efforts, Montreal faces challenges like traffic congestion, aging infrastructure, and balancing transportation demands with environmental sustainability goals. The city has set an endeavor to reduce greenhouse gas emissions, increase the use of public and active transportation modes, and enhance urban mobility systems' efficiency and accessibility. The REM is an example of the city and region's effort on automating and electrifying the mass transit systems.

As the present paper is being written, the city is in the process of producing its 2050 Land Use and Mobility Plan (Montreal's 2050 Land Use and Mobility Plan: https://montreal.ca/en/articles/creating-2050-land-use-and-mobility-plan-15575). The plan fully recognizes the interactions of land use and mobility services and, hence, simultaneously plans for the two major aspects. There are six phases in this process with Phase I (Assessing the city's situation) and Phase II (Imagining possible futures) having already been completed. When we are writing this paper, the city is in Phase III collecting public comments regarding the current city vision proposed in Phase II. The city will improve the vision and the implementation strategy (Phase IV), and in Phase V, the city council will provide additional comments and decide whether to adopt the plan. Phase VI will focus on implementing the plan and monitoring the progress (and adjusting when necessary). The process started in around 2019 (and just finished Phase II in the year 2024), indicating the heavy effort and cost of producing such a plan. The previous plan was released in 2008 with the following five main objectives:

- Create optimal conditions for getting around the city.
- Improve residents' quality of life.
- Improve the environment.

- Support the vitality of Montreal's economy.
- Integrate the planning of transportation with that of land use.

The SAEMS planning in the present case study has a narrower scope than that of the city's ongoing land use and mobility plan, as the case study focuses on planning the SAEMS (as a subsystem of Montreal's mobility system). However, the publicly available information released by the present planning process indeed provides useful information. In designing the prompts, we refer to the existing reports, such as the report presenting the results of citizen focus groups, the report on the city project ideation workshop, the sustainable mobility survey, and (Phase II's) *City Vision: Imagining the Montreal of 2050.* We leave the study of expanding the study scope using the proposed synthetic participatory approach to a broader transportation planning process for future research.

*4.2. Prompting and Output Processing*

Creating Human Agents and AI Assistants. The process begins with introducing the decision context and purpose to the LLM-enabled AI agent that will simulate the participatory planning process. For $\rho_0$, we have:

*"You are an expert in simulating a participatory planning process for the City of Montreal's future shared automated electric mobility systems (SAEMS). Suppose the planning activity occurs in the year 2024, and the planning horizon year is 2044."*

Then, we prompt digital avatars to be constructed representing Montreal's stakeholders, such as residents, policymakers, urban planners, and transportation engineers (including modelers). Instead of using the prompt to detail their socioeconomic backgrounds, temperaments, and preferences, this case study directly prompts a request for the generation of $S = \theta_1$ number of stakeholders. The profiles (and the associated probability distributions) of the generated stakeholder avatars are one main aspect of our research interest. For $\rho_1$, we have:

*"Create $\theta_1$ stakeholder(s) as their digital avatar(s)."*

Due to the stochasticity of LLM outputs, the LLM will respond by requesting additional details instead of directly generating digital representatives, on rare occasions. In simulations, we drop these cases, though it is likely that further prompting can help utilize these cases.

Team Formation and Synthetic Participatory Process. In this phase, we use two sequential prompts to guide the AI agent. We simplify the case by not introducing any real stakeholder participant in the participatory process. The team composed of (only) digital representatives of the stakeholders engages in a participatory process that includes identifying existing conditions, objectives, and performance metrics; then, the team collaboratively generates solution alternatives (design alternatives and implementation plans). Various participatory methods, such as the Synthetic Delphi approach and Synthetic focus groups, are explored to determine their impact on the planning outcomes.

For $\rho_2^1$, we have:

*"Concretely simulate the process and obtain the results from using a $\theta_2^1$-round Delphi method to let them collaboratively identify issues, objectives, performance metrics, and decision weights for each round. Then synthesize the information to form 5 issues, 5 objectives, 10 performance metrics (0–10 for each metric), and decision weights (sum up to 1.0). Be clear which performance metric is for which objective."*

For $\rho_2^2$, we have

*"Concretely simulate a $\theta_2^2$-min free-style brainstorming session to generate 3 mutually exclusive SAEMS alternatives with detailed specifications and the corresponding 20-year implementation plans with 4-year intervals (with a specific monetary amount for each interval) under a total budget of \$$\theta_2^3$ million CAD (Net Present Value)."*

Evaluation and Decision-Making. The synthetic participants collaboratively develop models to evaluate solution alternatives using the multi-criteria evaluation and decision framework. For $\rho_3$, we have:

> *"Evaluate the alternatives using the previously identified performance metrics—make best guess about the values about the variables and probabilities collaboratively. Compare the final scores and recommend the best alternative."*

*4.3. Parameterization, Baseline, and Scenario Settings*

The participatory process identifies objectives, performance metrics, and how key variables interact within specific SAEMS and the broader transportation systems. We vary parameters, such as team size, participatory methods, and associated durations and efforts, to study the impact of the resultant SAEMS plans for Montreal. This parameterization allows for a systematic analysis of how variations in the planning process influence the identification of problems, objectives, performance metrics, and the development and evaluation of alternatives for Montreal's SAEMS.

Key parameters include team size ($\theta_1$), number of Delphi rounds for identifying problems, objectives, and performance metrics ($\theta_2^1$), the duration of free-style discussion and ideation ($\theta_2^2$) in minutes, and the budget constraints ($\theta_2^3$) in million Canadian dollars (net present value, NPV). We set the base scenario as $[\theta_1, \theta_2^1, \theta_2^2, \theta_2^3] = [10, 3, 90, 100]$. In the next section, we will first present the results (including the stochasticity) of the base scenario. Then, we will explore the impact of perturbating the base scenario on the results of the synthetic participatory process. Note that in the case study, we use words such as Delphi and free-style brainstorming or discussions as parameters rather than simulating the specific conversations or dialogues. We leave the study of varying the resolution of conversations or dialogues to future research.

## 5. Simulation Results and Discussions

The simulation outcomes are discussed under two main subsections: base scenario results and sensitivity analysis of different scenarios based on various prompt settings. The simulations are performed using the OpenAI API version "gpt-4-turbo" in February and March 2024. The temperature is set as 1.0 (a median number in terms of the randomness of the LLM output). One complete round of simulation instances is provided in Appendix A.

*5.1. Baseline Instance*

In this subsection, we analyze the results of multiple runs of the baseline scenario, where each simulation generates a separate set of five stakeholders. Table 1 shows the 10 generated digital representatives of the stakeholders in SAEMS. Each avatar has a unique perspective and stake in the successful planning and implementation of SAEMS, demonstrating the necessity of a collaborative approach that recognizes and addresses the wide range of concerns and objectives. We can further request additional detailing for each avatar, and we leave the investigation of the optimal prompting fidelity for a given SAEMS decision context to future research.

**Table 1.** Ten generated digital representatives of stakeholders and the summaries of their profiles.

| Stakeholder | Profile |
| --- | --- |
| – City Planner | An experienced urban planner specializing in transportation infrastructure, with a focus on land-use densification and sustainable and efficient mobility solutions for marginalized communities. |
| – Transport Engineer | A professional with expertise in the design and implementation of transportation facilities such as signal timing, intersection channelization, and roadside arrangement. |
| – Policy Maker | A government official responsible for developing and enacting policies and investments that promote transportation systems that are in favor of their constituent. |
| – Environmental Advocate or Specialist | A representative from a non-profit organization dedicated to reducing carbon emissions and promoting environmentally friendly transportation options. |

| | |
|---|---|
| – Community Representative | A respected member of a local community group, representing the interests and concerns of residents, who will be directly impacted by the SAEMS. |
| – Technology Entrepreneur or Innovator | The founder of a startup company specializing in shared autonomous vehicle technology and electric mobility solutions. |
| – Academic Researcher | An academic or scientist conducting research on the impacts of shared automated electric mobility systems on urban environments. |
| – Local Business Owner or Representative | Represents the interests of local businesses, focusing on how SAEMS can support economic growth and accessibility. Interested in how SAEMS can impact local commerce, potentially increasing foot traffic but also raising concerns about congestion and parking. |
| – Public Transit Authority Representative | An official from Montreal's public transportation agency focused on integrating SAEMS with existing transit networks. |
| – Public Safety Official or Expert | Concentrates on the safety implications of SAEMS, including emergency response, accident prevention, and security measures. |

The stakeholder avatars collaboratively identified and prioritized objectives for the potential SAEMS in Montreal, highlighting the importance of sustainability and accessibility as the leading concerns. The group collaboratively determined the performance metrics for concretely evaluating these objectives, with decision weights reflecting the collective priorities of this diverse group. Table 2 shows an instance.

**Table 2.** Final List of Identified Issues, Objectives, Performance Metrics, and Decision Weights.

| Objective | Performance Metrics | Weight |
|---|---|---|
| Traffic flow improvement | Average travel time reduction, congestion reduction percentage, public transit integration level | 0.15 |
| Environmental footprint minimization | Greenhouse gas emission reduction, energy efficiency improvement, noise pollution reduction | 0.15 |
| Equity enhancement | Accessibility index for disadvantaged groups, affordability index for transportation | 0.10 |
| Infrastructure Integration | SAEMS network coverage, intermodal connectivity level | 0.30 |
| Technological innovation fostering | Adoption rate of new SAEMS features, collaboration level with tech companies | 0.30 |

The brainstorming session yielded three distinct SAEMS alternatives focusing on comprehensive transit solutions, modular mobility integration, and environmental sustainability.

**Alternative 1: The Eco-Friendly System** focuses on minimizing environmental impacts and promoting sustainability. (1) Deploy SAEV fleets with high energy efficiency and low emissions; (2) Install renewable energy charging stations powered by solar and wind energy; and (3) Promote the implementation of green roofs, permeable pavement, and urban green spaces to mitigate the heat island effect and enhance biodiversity. The collaboratively generated implementation plan is as follows:

- Years 1–4: Pilot phase in downtown areas, deploy 100 SAEVs and 50 charging stations (CAD 20 million).
- Years 5–8: Expand to suburban areas, increase fleet size to 300 SAEVs and 150 charging stations (CAD 30 million).
- Years 9–12: Integrate with public transit network, add 500 SAEVs and 250 charging stations (CAD 40 million).
- Years 13–16: Enhance infrastructure resilience, upgrade700 SAEVs and 350 charging stations (CAD 25 million).
- Years 17–20: Evaluate and improve system efficiency, optimize fleet operations, and add another 800 SAEVs and 400 charging stations (CAD 35 million).

**Alternative 2: The Equity and Accessibility System** focuses on improving transportation access for all communities. Specifications: (1) Inclusive Design: Implement accessible vehicle designs and infrastructure to accommodate individuals with disabilities; (2) Affordable Fares: Introduce fare structures that consider income levels and provide subsidies for low-income users; and (3) Accessible Stations: Upgrade existing and new stations to be accessible to individuals with mobility challenges. The collaborative generated implementation plan is as follows:

- Years 1–4: Accessibility upgrades in key locations, retrofit 50 stations and 200 vehicles (CAD 15 million).
- Years 5–8: Expand service to underserved areas, add 500 vehicles and 100 new stations (CAD 35 million).
- Years 9–12: Implement fare subsidies for low-income users, provide discounts and incentives (CAD 30 million).
- Years 13–16: Enhance accessibility features, add 300 vehicles and upgrade 50 stations (CAD 25 million).
- Years 17–20: Conduct community outreach and feedback, engage with stakeholders to improve accessibility (CAD 25 million).

**Alternative 3: The Technology-Promoting System** focuses on pioneering promising technologies and enhancing traveler experience. Specifications: (1) Deploy a fleet of SAEVs with advanced safety features and AI-driven capabilities; (2) Implement AI algorithms to optimize vehicle routes and minimize congestion; and (3) Upgrade infrastructure to support smart connectivity of civil infrastructure and facilities, enabling real-time communication between vehicles and infrastructure. The collaborative generated implementation plan is as follows:

- Years 1–4: Pilot SAEV fleet, introduce 50 SAEVs and test AI algorithms (CAD 25 million).
- Years 5–8: Expand the SAEV fleet and implement AI-driven real-time route optimization, add 200 SAEVs and optimize dedicated AV routes (CAD 40 million).
- Years 9–12: Upgrade infrastructure for smart connectivity, install sensors and communication systems (CAD 30 million) (to further improve the cybernetic capabilities of Montreal's SAEMS.)
- Years 13–16: Enhance user experience with app integration, develop user-friendly apps for trip planning and payment (CAD 20 million).
- Years 17–20: Evaluate and integrate emerging technologies, upgrade systems based on public feedback and technological advancements (CAD 35 million).

The descriptions of the three alternatives are summarized in Table 3. Figure 4 illustrates the corresponding investment profiles. Each alternative comes with a pragmatic 20-year phased implementation plan (with 4-year intervals), indicating financial implications to the stakeholders.

Table 4 shows the results of the collaborative multi-criteria evaluation of the three alternatives. Based on the provided metrics and estimated scores, the "Eco" alternative emerges as the most favorable option in terms of both the total score (38) and the weighted score (7.75). This alternative is scored high in integration with emission minimization and its complementary nature to infrastructure such as public transit, factors that are critical to the successful and sustainable implementation of SAEMS. Although the "Equi" and "Techno" alternatives present compelling benefits in areas such as congestion reduction, accessibility, and affordability, they fall slightly behind the "Eco" alternative in terms of comprehensive fit, innovation, and public transit synergy. The "Eco" alternative also has a moderate requirement for financial commitment in the initialization phase (the first four years), allowing for lower risk to the public agency (compared to the "Techno" alternative). Continued stakeholder engagement and detailed feasibility studies will be essential in refining and implementing this ambitious plan.

**Table 3.** Final List of Design Alternatives, Corresponding Implementation Plans.

| Alternative | 2024–2028 | 2028–2032 | 2032–2036 | 2036–2040 | 2040–2044 |
|---|---|---|---|---|---|
| **“Eco-SAEMS”:** Focus on minimizing environmental impact and promote sustainability. Specifically, deploy SAEVs, renewable energy charging stations, green infrastructure. | | | | | |
| | Pilot phase in downtown areas (CAD 20 million) | Expand to suburban areas (CAD 30 million) | Integrate with public transit network (CAD 40 million) | Enhance infrastructure resilience (CAD 25 million) | Evaluate and improve system efficiency (CAD 35 million) |
| **“Equitable-SAEMS”:** Improve transportation access for all communities. Specifically, inclusive design, affordable fares, accessible stations. | | | | | |
| | Accessibility upgrades in key locations (CAD 15 million) | Expand service to underserved areas (CAD 35 million) | Implement fare subsidies for low-income users (CAD 30 million) | Enhance accessibility features (CAD 25 million) | Conduct community outreach and feedback (CAD 25 million). |
| **“Techno-SAEMS”:** Pioneer new technologies and enhance user experience. Specifically, deploy SAEVs, AI-driven route optimization, smart infrastructure. | | | | | |
| | Pilot autonomous vehicle fleet in dedicated areas or zones (CAD 25 million). | Expand autonomous fleet and implement AI-driven route optimization (CAD 40 million) | Upgrade infrastructure for smart connectivity (CAD 30 million) | Enhance user experience with app integration (CAD 20 million) | Evaluate and integrate emerging technologies (CAD 35 million). |

[Source: Own Elaboration.]

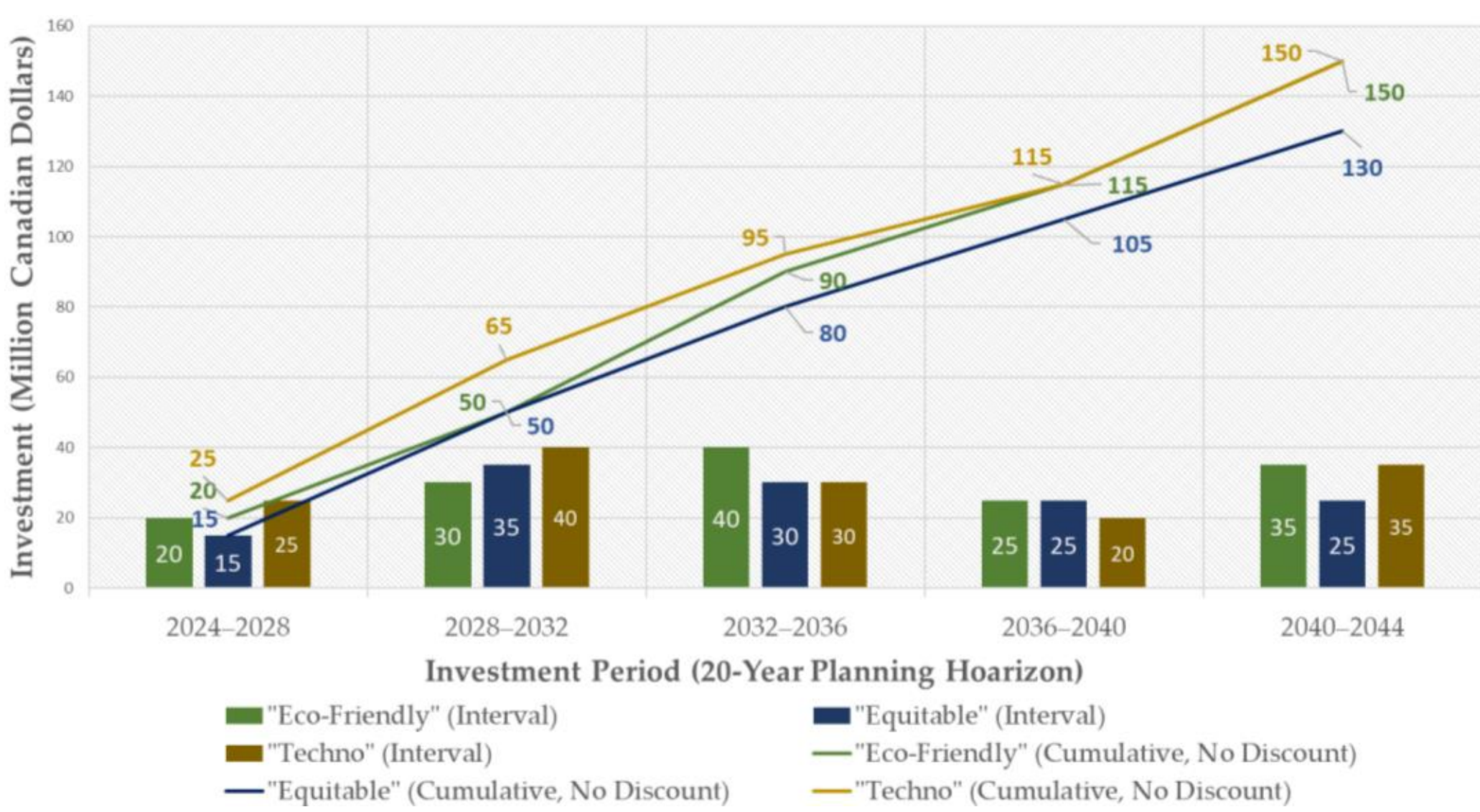


**Figure 4.** Period and cumulative investment for the three SAEMS alternatives. [Source: Own Elaboration.]

**Table 4.** Evaluation Scores (By Metrics) and the Total Weighted Scores for Each Alternative.

| Objectives | Weights | “Eco” | “Equi” | “Techno” |
|---|---|---|---|---|
| Traffic Flow Improvement | 0.15 | 7 (1.05) | 5 (0.75) | 9 (1.35) |
| Environmental Footprint Minimization | 0.15 | 8 (1.20) | 6 (0.90) | 7 (1.05) |
| Equity Enhancement | 0.10 | 7 (0.70) | 9 (0.90) | 4 (0.40) |
| Infrastructure Integration | 0.30 | 9 (2.10) | 5 (1.50) | 8 (2.70) |
| Technological Innovation Fostering | 0.30 | 7 (7.75) | 5 (1.50) | 8 (2.40) |
| **Total Score (Avg Weighted Score)** | 1.00 | **38 (7.75)** | 30 (5.55) | 36 (7.60) |

### *5.2. Sensitivity Analysis*

This subsection explores how the results of the synthetic participatory planning process change in response to different participation settings. The analysis aims to understand the impact of variations in team

size, participatory methods, and the duration/effort allocated to different phases of the process on the planning outcomes.

5.2.1. Impact on Stakeholder Profiles

To analyze the synthetic stakeholder profiles generated across multiple simulations, we employ a horizontal bar chart for different group sizes, where we visualize the frequency of each stakeholder type. This provides a clear view of which profiles are most commonly generated and how they change in response to the change in team size. Figure 5 shows the distribution of a stakeholder type to appear at least once from 150 simulations for each team size setting. Table 5 shows summaries of the profiles of the generated stakeholders through simulations.

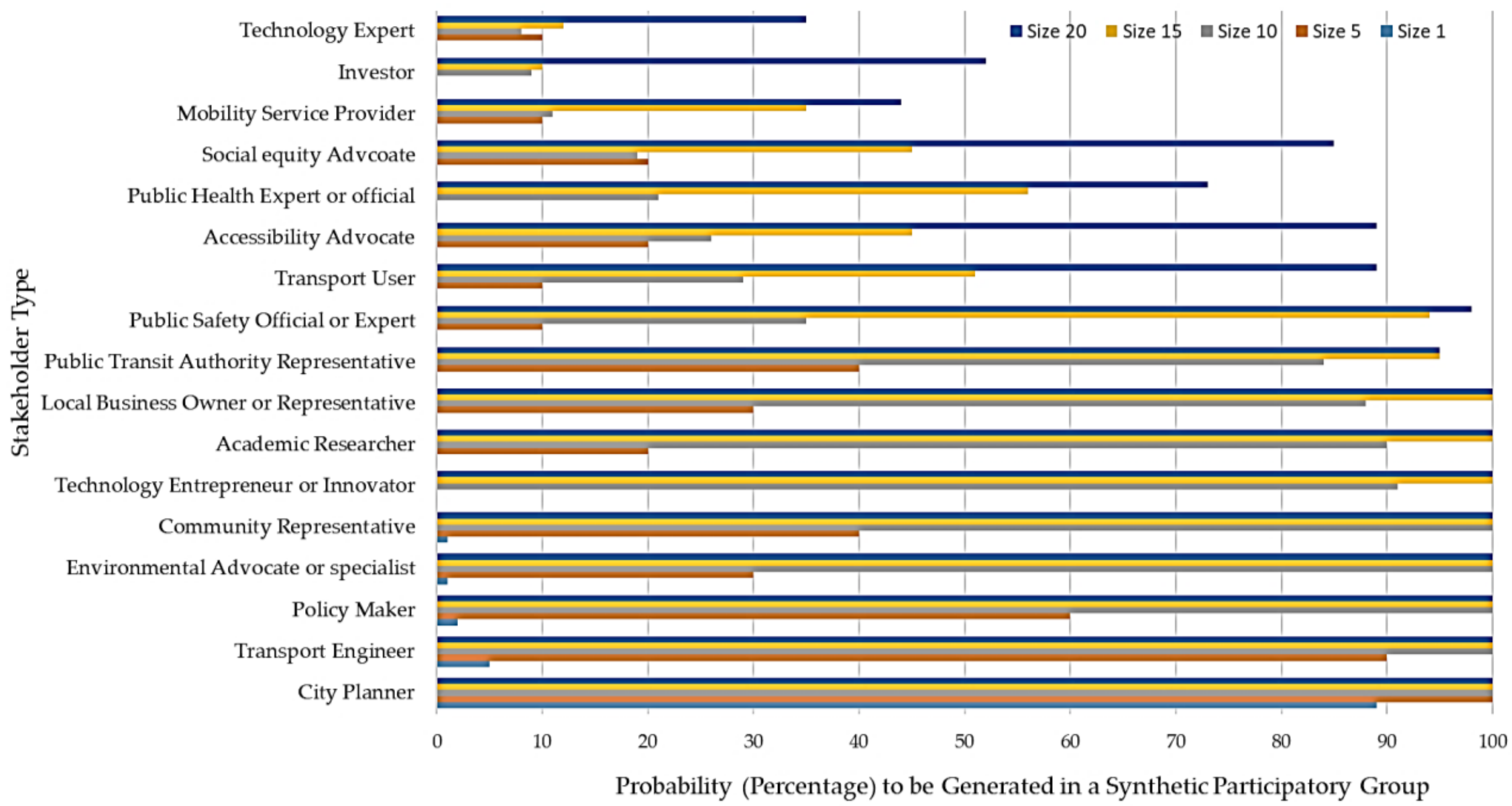


**Figure 5.** Frequency of different stakeholder profiles generated across simulations.

With the increase in the number of stakeholders, generated synthetic stakeholders start having more sub-categories of profiles. For example, one simulation might generate a resident's association leader, a senior citizen representative, and a youth representative, all of whom fall into community representatives. For another example, the LLM starts generating different types of planners that focus on land zoning, public parks, transportation, etc. It is also important to note that with the increase from 1 to 10 stakeholders, the coverage of stakeholders (at least by one stakeholder) starts to increase, while from 10 to 20, most additional stakeholders fall into existing categories (i.e., it becomes more likely to have more than one stakeholder in one category). This phenomenon implies the potential existence of an optimal team size and shows that explicitly requesting the LLM to generate more than one synthetic participant can help avoid an implicit assumption about who the default stakeholder is.

5.2.2. Impact on Objectives and Performance Metrics

The influence of varying team sizes and the number of Delphi rounds on the articulation of objectives and performance metrics, including the assignment of decision weights, is critically evaluated through comparative analysis across multiple simulation scenarios. Simulations incorporating different numbers of stakeholder avatars are conducted to discern the effects of larger team compositions. While more extensive teams tend to offer a wider array of perspectives, they also introduce increased complexity in interactions and additional challenges for reaching a consensus. To systematically analyze these dynamics, the

objectives identified across simulations are categorized into ten distinct groups, as detailed in Table 6. Figure 6 presents a summary of the average decision weights assigned in each configuration along with their standard deviations (the total length of an error bar represents two standard deviations ($2\sigma$)), reflecting the inherent stochasticity of the simulation outcomes. To simplify the notations, we denote a scenario with a team size of 10 and the Delphi round number 3 as "S10R3." A similar principle applies to other scenario settings.

**Table 5.** Summaries of example profiles for the generated stakeholders.

| Stakeholder | Example Profile |
|---|---|
| – City Planner | An experienced urban planner specializing in transportation infrastructure, with a focus on sustainable and efficient mobility solutions. |
| – Transport Engineer | A professional with expertise in the design and implementation of advanced transportation systems, including automated and electric vehicles. |
| – Policy Maker | A government official responsible for developing and enacting policies that support the integration of SAEMS into the city's transportation network. |
| – Environmental Advocate or Specialist | A representative from a non-profit organization dedicated to reducing carbon emissions and promoting environmentally friendly transportation options. |
| – Community Representative | A respected member of a local community group, representing the interests and concerns of residents, who will be directly impacted by the SAEMS. |
| – Technology Entrepreneur or Innovator | The founder of a startup company specializing in autonomous vehicle technology or electric mobility solutions. |
| – Academic Researcher | An academic or scientist conducting research on the impacts of shared automated electric mobility systems on urban environments. |
| – Local Business Owner or Representative (e.g., Restaurant, Tourism/Hotel, Real Estate) | Represents the interests of local businesses, focusing on how SAEMS can support economic growth and accessibility. Interested in how SAEMS can impact local commerce, potentially increasing foot traffic but also raising concerns about congestion and parking. |
| – Public Transit Authority Representative | An official from Montreal's public transportation agency focused on integrating SAEMS with existing transit networks. |
| – Public Safety Official or Expert | Concentrates on the safety implications of SAEMS, including emergency response, accident prevention, and security measures. |
| – Transport User | A regular commuter who relies on public transportation. This person is interested in how SAEMS can improve their daily commute and provide more flexible travel options. They are concerned about the affordability and reliability of SAEMS compared to traditional public transit. |
| – Accessibility Advocate | A disability rights activist working to ensure that SAEMS are accessible to all, including people with disabilities and seniors. This person emphasizes the importance of universal design and user-friendly interfaces. |
| – Public Health Expert or official | Examines the health implications of SAEMS, including air quality, noise pollution, and active transportation options. |
| – Social Equity Advocate | A representative from a social justice organization. SEA focuses on ensuring that SAEMS are accessible and affordable to all segments of the population, particularly marginalized communities, and that the planning process is inclusive and participatory. |
| – Mobility Service Provider | The CEO of a company that offers shared mobility services, such as car-sharing or ride-hailing. This person is also considering expanding the market to food delivery services. |
| – Investor | A financial stakeholder interested in funding innovative transportation projects, particularly those involving automation and electrification. |
| – Technology Expert | Specializes in autonomous vehicle technology and AI systems. Provides insights into the technical feasibility and innovation potential of SAEMS. |
| – Traffic Management Specialist | With expertise or experience in managing urban traffic flows and parking, this avatar is concerned with how SAEMS will impact congestion and traffic patterns. They work on developing strategies to optimize traffic management and reduce travel times. |

**Table 6.** Summary of commonly identified alternatives during participatory planning. The objectives identified across simulations are categorized into ten distinct groups.

| Acronym | Objectives | Generated Performance Metric Examples |
|---|---|---|
| ENRM | Environmental Impact and Sustainability | $CO_2$ emissions reduction, energy efficiency |
| EFFIC | Congestion Reduction and Mobility Efficiency | Average Travel Time, Traffic Flow Improvement, vehicle kilometers traveled |
| SAFTY | Safety Enhancement, Regulatory Compliance | Number of accidents, comfort, and perceived safety |
| ACCESS | Accessibility Improvement | Coverage Area of SAEMS, service availability, number of hospitals, and libraries in 15 min buffer |
| PubTrans | Integration with public transit and other public services | Increase in transit coverage, multimodal transit usage, transit connectivity |
| Econ | Economic Viability | Cost-effectiveness, return on investment, employment stimulation |
| Pub-Adopt | Maximize public support and acceptance, market penetration | Public positive perception or acceptance rate. Potential user adoption rate, market share, user rating |
| R&R | Technology and infrastructure integration and reliability, cost-effectiveness, Security, Resilience | System uptime, passenger feedback, tech innovation rate. Resilience to natural disasters, potential frequency, and severity of cyberattacks, emergency response time |
| Equity | Equity and Inclusivity | Service Accessibility for disadvantaged groups, fair pricing, affordability |
| Infra Use | Efficient energy and infrastructure utilization | Existing facility and equipment utilization. Including promotion of urban green space, even charging station usage |

As evidenced across various scenarios, the objectives concerning environmental and sustainability impact, congestion, safety, and accessibility, consistently received higher decision weights. Interestingly, as the team size increases, the distribution of weights becomes more even, suggesting that the weights of less common concerns or objectives will be more important among a larger group of stakeholders. However, the impact of increasing the number of Delphi rounds appears nonlinear. Transitioning from one to three stakeholders tends to produce more specific objectives and lower variation in decision weights, possibly indicating the LLM's capacity for simulating a consensus-building effect. Conversely, expanding the team from three to five stakeholders does not yield significant changes, suggesting diminishing sensitivity to additional Delphi rounds. Larger groups may also require more rounds to achieve consensus, which seems consistent with the common experience.

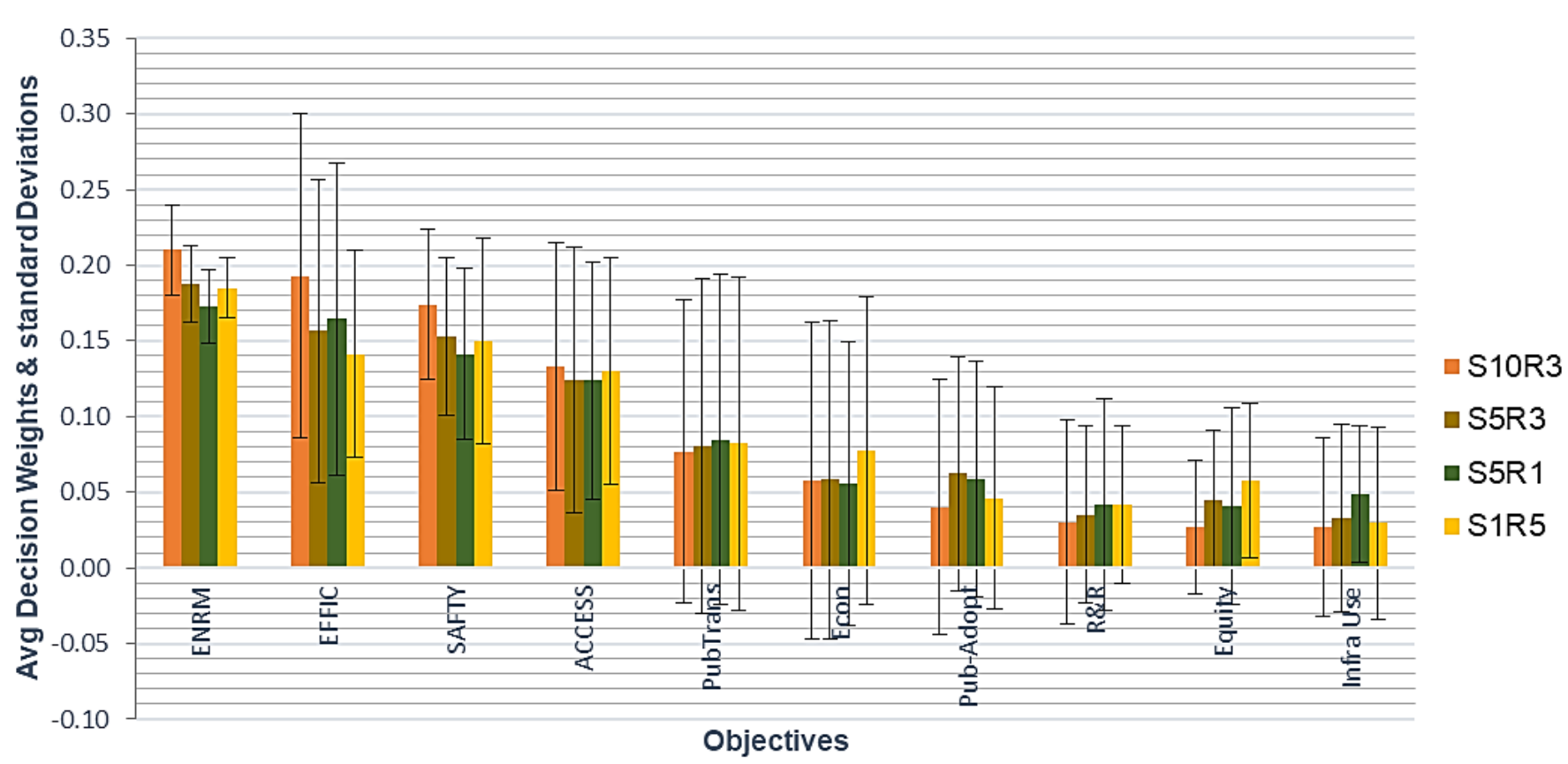


**Figure 6.** Decision weights for each category of objective and for each of the four parameter settings (team size and Delphi number of rounds).

Moreover, the error bars that enter the 'negative' conditions suggest that when decision weights are low, they tend to be highly skewed. Specifically, certain alternatives are seldom selected (thus, often receiving zero weight); however, when these alternatives are chosen, their weights typically range from 0.10 to 0.20 rather than much lower (but nonzero) weights. This pattern underscores the importance of framing and participatory methods in replicating results for validation and mitigating common biases in participatory decision-making.

Further analysis suggests that increasing the number of Delphi rounds beyond three yields minimal impact on consensus quality. Both small groups with fewer rounds and larger groups with more rounds tend to lead to more evenly distributed weights, posing an interesting question on whether such phenomena also occur in real participatory processes. The exploration of a more diverse range of participatory methods—transitioning from structured approaches like the Delphi method to more free-form discussions—could provide insightful contrasts. Different methods likely influence the level of consensus, depth of discussion, and creativity of the solutions generated, which, in turn, influence the resultant plan.

#### 5.2.3. Impact on Design Alternatives and the Evaluation

This subsection conducts a sensitivity analysis to evaluate the impact of team size, the duration of free-style ideation sessions, and budget constraints on the identification of issues and the development of alternatives within the synthetic participatory process. We hypothesize that longer durations might allow for more in-depth discussions, while shorter sessions could necessitate more efficient and focused deliberations. Our results help assess the robustness of baseline scenario outcomes and identify potential areas for optimization. Settings that consistently yield better alignment with planning objectives might be favored in future simulations. We explore the diversity of design alternatives conceived by synthetic teams and their alignment with the identified objectives and performance metrics. The analysis focuses on several common design alternatives generated during the planning process, as summarized in Table 7. The frequency and distribution of these alternatives across simulations highlights trends and preferences among synthetic teams. For instance, a frequent selection of a particular alternative might indicate a strong emphasis on optimizing certain aspects, like travel efficiency.

Each design alternative undergoes evaluation based on its alignment with established objectives and performance metrics. For example, an alternative might score highly on environmental impacts due to the integration of electric transit systems, while another might be favored for enhancing accessibility.

The variability in the evaluation results is analyzed through the distribution and scores for each performance metric. A radar chart (Figure 7) is utilized to visualize the variability in priorities among synthetic teams across simulations. This method allows us to identify common trends or outliers in planning outcomes. "S$\theta_1$-R$\theta_2^1$-D$\theta_2^2$-B$\theta_2^3$" represents team size $\theta_1$, $\theta_2^1$ number of Delphi rounds, $\theta_2^2$-minute free-style discussion/brainstorming, and \$$\theta_2^3$ million Canadian dollars budget (NPV). We select the most representative cases for analysis instead of averaging for each scenario setting to avoid obscuring trade-offs inherent in the planning process.

**Table 7.** Common design alternatives generated during the planning process.

| | Alternative | Example Description |
|---|---|---|
| 1 | Advanced/multimodal mobility network | Fully integrate SAEMS with public transit, bike sharing, offering seamless multimodal trip planning and fare integration. |
| 2 | Integrated MaaS platform | Developing an integrated Mobility as a Service (MaaS) platform that offers a seamless and personalized mobility experience for users, integrating various modes of transportation including SAEMS. |
| 3 | Urban mobility hub | Focuses on creating urban mobility hubs that serve as centralized locations for various modes of transportation, including SAEMS, public transit, cycling, and walking. These hubs are designed to improve connectivity and accessibility while reducing the reliance on private vehicles with sufficient charging (and discharging) capabilities. |
| 4 | Automated shuttles | A network of automated shuttles providing first-mile/last-mile connectivity. |
| 5 | Basic System Upgrade | Implement a basic SAEMS with limited coverage and vehicle fleet size. |
| 6 | Sustainable Urban Mobility | Integrated SAEMS with green spaces and urban planning to create a more sustainable and livable city environment. |
| 7 | Smart transport infrastructure | Focuses on upgrading the city's road infrastructure with smart sensors and communication technologies to improve traffic management and enhance the efficiency of SAEMS. Use of AI. |
| 8 | Urban Air Mobility System | Implement a network of autonomous aerial vehicles for passenger and cargo transport, reducing ground congestion and improving accessibility. |

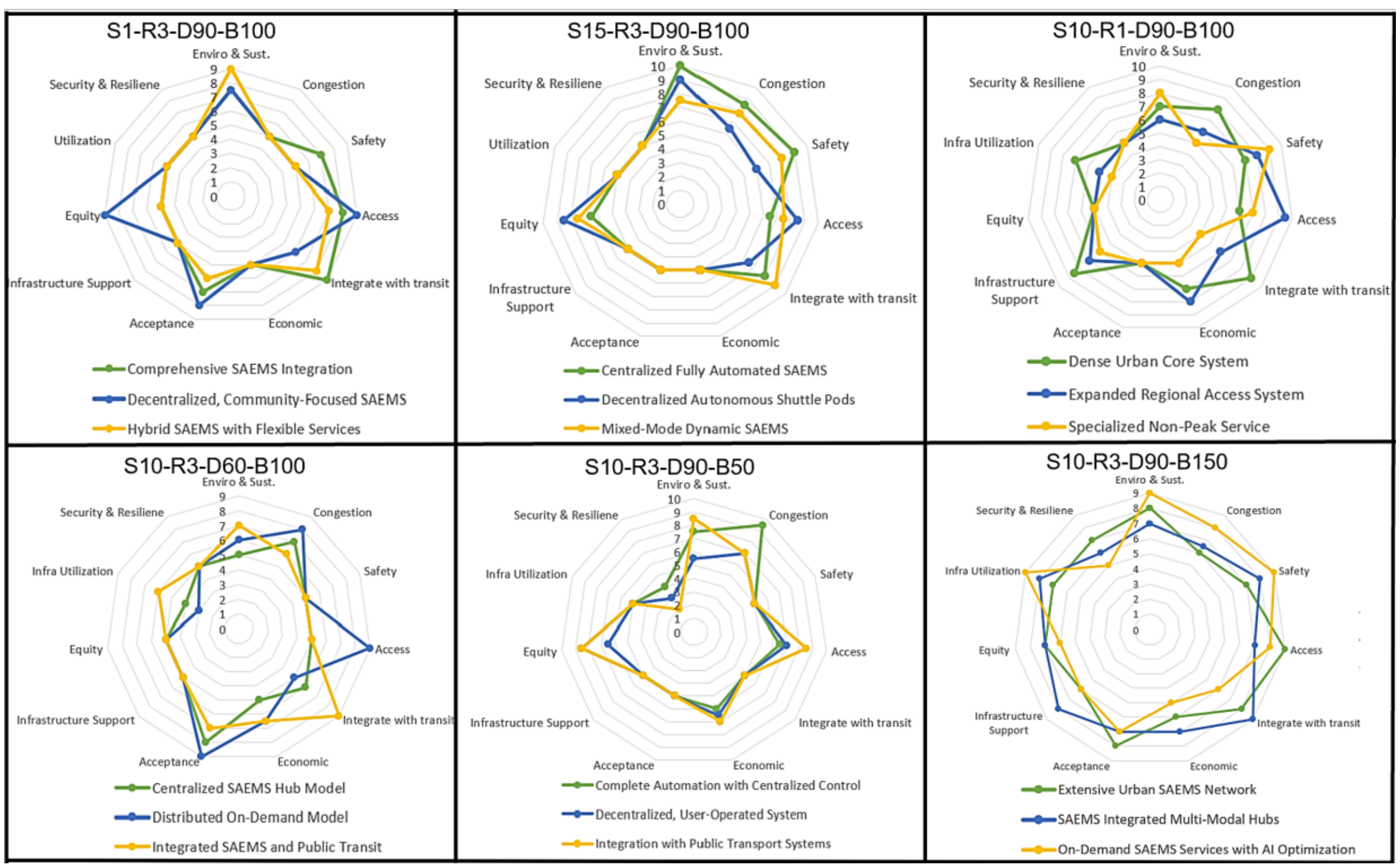

**Figure 7.** Radar chart comparison among the representative alternatives generated from six different settings.

Smaller teams, exemplified by the scenario "S1-R3-D90-B100", tend to favor alternatives that align closely with the expertise of urban planners, as when the team size is one, an urban planner profile is the most likely one to be generated. Larger teams (15 members or more), exemplified by the scenario "S15-R3-D90-B100", do not exhibit significant changes in the evaluation, suggesting a saturation point in the influence of team size on decision-making. The duration of free-style discussions for conceiving solution alternatives shows that longer durations tend to generate alternatives that balance performance metrics, as exemplified by the scenarios “S10-R3-D90-B100” and “S10-R3-D60-B100”. Additionally, it is evident that budget constraint is an important factor. Manifested by the greater area of the “S10-R3-D90-B150,” the trade-off effect among different objectives is clearly lessened thanks to the higher available funding. We leave a more comprehensive analysis of the interactions among parameters for future research.

## 7. Conclusions

This study highlights a synthetic participatory approach that utilizes LLMs to create digital avatars representing various stakeholders in urban mobility planning. Our case study in Montreal demonstrates the effectiveness of this method in addressing the research questions posed in Section 1. First, we explored how digital avatars, calibrated with data from real stakeholders, enhance the inclusivity and effectiveness of the transportation planning process. Our findings indicate that digital avatars significantly broaden stakeholder engagement, overcoming logistical and financial constraints associated with conventional methods. This inclusivity ensures that a diverse range of perspectives is considered, leading to more comprehensive and representative urban mobility solutions. Second, we examined how the integration of digital and real stakeholders provides a scalable and adaptable framework for planning shared automated electric mobility systems. The structured and parameterized workflow we developed allows for the simulation of diverse stakeholder interactions and scenarios, making the planning process more adaptable to various urban contexts and planning scenarios. This adaptability is particularly beneficial in complex urban settings like Montreal, where the mobility challenges necessitate nuanced and flexible approaches. Third, we investigated how our method improves the interpretability and controllability of planning outcomes compared to traditional methods. By leveraging digital avatars and a systematic workflow, the approach enhances the transparency and traceability of decision-making processes. This structured approach not only enriches the exploration of potential outcomes and strategies but also ensures that urban mobility plans are resilient and forward-looking.

The implications of this research are significant for human-centered urban planning and transportation system engineering. The use of generative AI, particularly LLMs, facilitates the assimilation of volumes of unstructured data and the simulation of complex stakeholder dynamics, which are challenging to test and manage in real settings. This is particularly beneficial in diverse urban settings like Montreal, where the complexity of mobility challenges necessitates nuanced understanding and approaches. While the approach introduces new challenges, such as ensuring the accuracy and authenticity of digital avatar representations and managing the complexity of AI systems, the potential benefits suggest a promising direction for future urban mobility planning. As LLMs continue to evolve and prompting techniques are refined, we hope this paper invites more researchers to investigate this promising area of LLM applications in transportation planning.